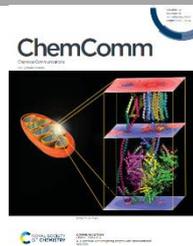

## COMMUNICATION

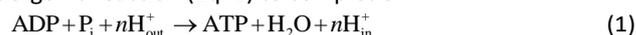

# ATP Synthase: A Moonlighting Enzyme with Unprecedented Functions

Jean-Nicolas Vigneau,[a,b] Peyman Fahimi,[a,c] Maximilian Ebert,[d] Youji Cheng,[e] Connor Tannahill,[f] Paul Muir,[f] Thanh-Tung Nguyen-Dang,[a] and Chérif F. Matta*[a,c,e,g]

**ATP synthase's intrinsic molecular electrostatic potential (MESP) adds constructively to, and hence reinforces, the chemiosmotic voltage. This ATP synthase voltage represents a new free energy term that appears to have been overlooked. This term is at least roughly equal in order of magnitude and opposite in sign to the energy needed to be dissipated as a Maxwell's demon (Landauer principle).**

Since Jacopo Tomasi introduced the study of the molecular electrostatic potential (MESP) decades ago,[1-5] molecular electrostatics have been implicated in enzyme catalysis,[6-8] and in predicting the response of molecules to external electric fields.[9;10] This state of affairs has prompted the use of high resolution X-ray diffraction to determine the electric field within enzyme active sites,[11;12] the use of small reporter molecules (*e.g.* CO, HCN) to determine the change in the local field accompanying site-directed mutagenesis,[13;14] and the use of the vibrational Stark effect[15] to establish the nature of exotic ions.[16] Some authors solve the Poisson-Boltzmann equation numerically to explore the electric potentials within the ribosomal tunnel[17] or within voltage-dependent channels,[18] to name a couple of examples.

In this paper ATP synthase (ATPase) is shown to play a role over and above its putative role as an enzyme. While this enzyme exists in other environments than mitochondria, that is, in bacteria and in chloroplasts, the essential physics is the same and the discussion here is centred on this organelle.

The mitochondrion is the "*powerhouse of the cell*"[19] being the site for the Krebs cycle, the electron transport chain (ETC), and the oxidative phosphorylation. The inner mitochondrial membrane (IMM) is where the exergonic reactions of the ETC take place. The $\Delta G$ released from the ETC is coupled with pumping $H^+$ against the concentration- and electrical-gradients from the mitochondrial matrix (in) to the inter-membrane space (out). The circuit is completed when protons return to the mitochondrial matrix through ATP synthase bringing the endergonic reaction (Eq. 1) to completion:

$$ADP + P_i + nH_{out}^+ \rightarrow ATP + H_2O + nH_{in}^+ \quad (1)$$

where $n \approx 3$ protons translocated/ATP.

Following Johnson and Knudsen (J&K)[20-22] who resolved the (then) long-standing paradox of the kidney's thermodynamic inefficiency,[23] it has been shown that ATP synthase must dissipate entropy to keep functioning.[24-26]

The inefficiency of the kidney hinges on equating "useful work" with osmotic work (only) - ignoring its regulatory function as a "selector" of ions ($Na^+/K^+$).[20-22] In selecting ions, the kidney is a realization of "*Maxwell's demon*" as J&K realized more than half a century ago. The demon must dissipate a minimum of $k_B \ln 2$ per bit of "erased information" (*Landauer limit*).[27-29] This is non-negotiable; it is a law of nature just like the Heisenberg indeterminacy principle, and is mechanism-independent. This is one of the reasons a hard drive heats up when one deletes large amounts of data. In the case of ATP synthase, to recognize one proton from the background heat-noise, at 98% fidelity (pH uncertainty in the gap is $\approx$ 2-3 %), the actual cost of a recognition event is $\approx 3k_B T$,[30] which at room temperature is around 10 kJ/mol.

We were not the first to realize that molecular machines such as ATP synthase are embodiments of Maxwell's demon,[31;32] but we were the first to emulate the calculations of J&K on this enzyme.[24-26] As a result it has been proposed to revise the textbook 55% efficiency of ATP synthase to *ca.* 90%, bringing it closer to the 100% efficiency of its rotor-stator mechanism.[33]

Assuming one glucose unit produces $\approx$ 32 ATPs, two of which are produced in glycolysis and two in the Krebs Cycle, leaves $\approx$ 28 ATP produced by oxidative phosphorylation. Each ATP requires, on average, the translocation of about 3 protons. Hence, to keep operating, ATP synthase must dissipate $\approx 28 \times 3 \times 10 \approx$ 800-900 kJ/mol of glucose consumed.

[a.] Dép. de chimie, Université Laval, Québec, QC, Canada G1V0A6.
[b.] Institut des Sciences Moléculaires d'Orsay (ISMO), Université Paris-Saclay, 91400 Orsay, France.
[c.] Dept. of Chemistry & Physics, Mount Saint Vincent University, Halifax, NS, Canada B3M2J6.
[d.] Chemical Computing Group (CCG), Sherbrooke Street West, Montreal, QC, Canada H3A2R7.
[e.] Dept. of Chemistry, Saint Mary's University, Halifax, NS, Canada B3H3C3.
[f.] Dept. of Mathematics & Computing Science, Saint Mary's University, Halifax, NS, Canada B3H3C3.
[g.] Dept. of Chemistry, Dalhousie University, Halifax, NS, Canada B3H4J3.
† Electronic supplementary information (ESI) available: Steps for protein preparation, conditions/parameters of the calculations, and a description (in addition to a complete listing) of the in-house Python 3.9.0 code used to manipulate the data, reorient the proteins, and analyze the MESPs – are all included in one Word document.

Because the Gibbs energy released from H⁺ translocation depends on local conditions, the conversions of glucose into a H⁺ gradient and eventually into ATP are not stoichiometric ($n \approx$ (not exactly = 3), in Eq. 1). Therefore, it is not difficult to miss $3k_BT$/proton $\approx$ 10 kJ/mol (37°C). *But has nature devised a workaround for ATP synthase to pay this "mandatory" tax to stay alive? It turns out that the answer is "yes"*. ATP synthase's very structure is associated with a MESP that adds constructively to the chemiosmotic voltage and, hence, contributes an additional $\Delta G$ term of the same order of magnitude (at least) and opposite in sign.

High-resolution X-ray diffraction- and electron microscopy (EM)-based structures of ATP synthase from five different species were obtained from the literature. These include structures from bacteria [*Paracoccus denitrificans* (PDB# 5DN6),[34] and *Bacillus sp.* (strain PS3) (PDB# 6N2Y)[35]], two fungi [yeast *Saccharomyces cerevisiae* (PDB# 6CP6),[36] and *Yarrowia lipolytica* [37](PDB# 5FL7)], and a mammal [wild boar (*Sus scrofa* (PDB# 6J5I))].[38] The structures were uploaded on the PDB2PQR and Adaptive Poisson-Boltzmann Solver (APBS) server[39] to calculate their respective MESPs after re-orienting every structure by aligning its long axis with the Cartesian *z*-axis. The technical details of the procedures are given in the **ESI‡**. The **ESI‡** also provides the full listing of an in-house Python utility code used in both the set-up of the calculation and the analysis and plotting of the results.

In all calculations, the pH was set to 7.0, the monovalent ion/counterion concentrations were taken to be 150 mM,[40] similar to those measured experimentally in the intermembrane gap using fluorescein-BSA,[41] the protein interior dielectric constant was set to 6, the solvent dielectric constant was set to 78.5, and all calculations were run at room temperature 298 K (25 °C).[40;42-44] The discretized Poisson-Boltzmann equation was then solved using AMBER[45-47] force-field atomic point charges to approximate the charge density of the protein. Since the results of all five structures are similar (details to be published elsewhere), here two examples are taken as the basis for most of the discussion, that is, *Yarrowia lipolytica* (fungus) (PDB # 5FL7) and *Sus scrofa* (wild boar) (PDB# 6J5I).

Fig. 1 displays the MESP of ATP synthase averaged over planes perpendicular to the principal (long) molecular axis (*z*-axis). To avoid artefacts due to the point charge representation of the protein charge density, the in-plane averaging of the MESP has been confined within 5 to 15 Å from the protein's surface. From the figure one can draw three observations (that occur in all studied five structures (as revealed in Fig. 2)): (*i*) There is a non-zero potential difference between the point of entry of the proton from the inter-membrane space into the F$_O$ unit and the proton's point of exit at the matrix side;[48] (*ii*) the *direction* of this ATP synthase-dependent voltage ($\Delta\Psi_{ATP\ synth.}$) is of the same sign as the charge-gradient-dependent IMM voltage ($\Delta\Psi$) and hence the two add-up constructively; (*iii*) there is a noted spike in positive potential ($\Delta\Psi^{\ddagger}_{ATP\ synth.}$) on entry and which constitutes an *activation barrier* regulating the rate of H⁺ translocation through the channel of the F$_O$ unit. This spike gives rise to a kinetic bottleneck ($\Delta G^{\ddagger}_{ATP\ synth.}$) never reported in the literature to our knowledge. As can be seen from Fig. 2, the qualitative general picture that emerges is consistent from one structure to another albeit with quantitative differences.

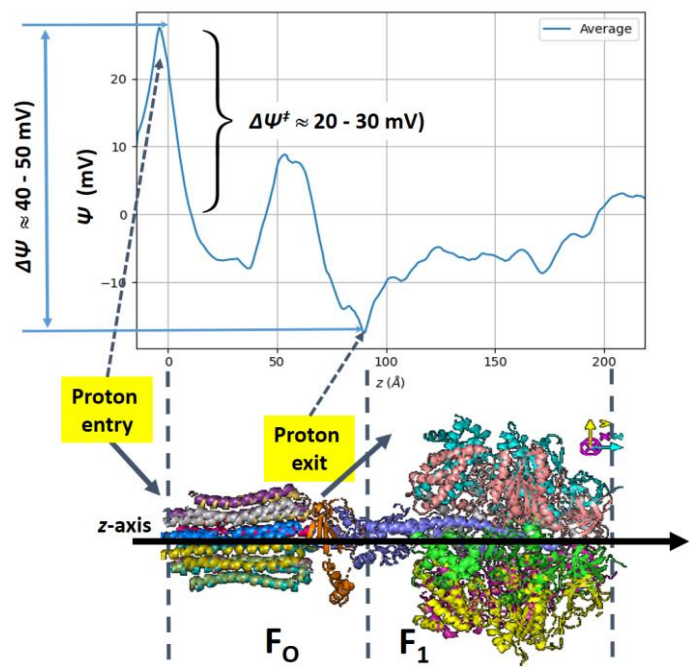

**Fig. 1** Calculated MESP of ATP synthase, $\Delta\Psi_{ATP\ synth.}$, averaged over each plane perpendicular to the *z*-axis between 5 and 15 Å from the protein (PDB # 5FL7 - *Yarrowia lipolytica* (fungus)) showing $\Delta\Psi_{ATP\ synth.}$, and $\Delta\Psi^{\ddagger}_{ATP\ synth.}$.

From point (*i*) and (*ii*) above, there is a Gibbs energy term of the same sign as that of the IMM charge gradient term that any proton will experience upon translocation through ATP synthase. The energy per mole of protons, solely due to the ATP synthase voltage difference, is $\approx$ 5 kJ/mol which is of the same order of magnitude as the $\approx$ 10 kJ/mol that need to be dissipated as a Maxwell's demon (*vide supra*).

Table 1 shows that the translocation of a mole of protons through the F$_O$ unit is accompanied with $\approx$ 1-5 kJ which originates from the intrinsic electric field (**E**) of the protein. Furthermore, the field creates a kinetic barrier regulating the entry of the protons into the F$_O$ unit of $\approx$ 1-3 kJ/mol. It is emphasized that these values are *averaged* over a full circle centred around the *z*-axis in any given plane. In this work, the entire protein is surrounded by a medium with the dielectric constant of water ~ 80 (see **ESI‡**). No attempt has been made to embed the F$_o$ unit into a lipid bilayer membrane (with a dielectric of around 2.0 – 2.5).[49] It is not possible to quantify the effect of such embedding on the potential differences without explicit calculations. What is predicted is that this drastic reduction in the dielectric constant, even for only a slab representing the membrane, will be accompanied with (possibly significantly) higher potential differences than those listed in Table 1. The values listed in the table can, hence, be regarded as *lower bounds*.

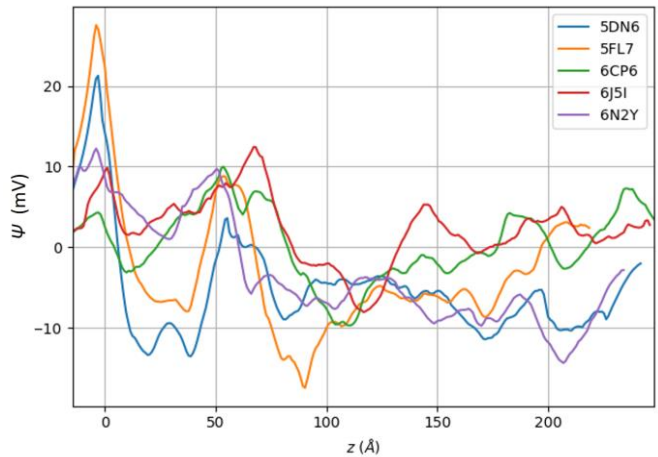

**Fig. 2** Calculated intrinsic MESP of ATP synthase, $\Psi_{\text{ATP synth.}}$, averaged over planes perpendicular to the *z*-axis between 5 and 15 Å from the protein for the five studied structures.

It is suggested, therefore, to append two new terms to the central equation of Mitchell's chemiosmotic theory to read:

$$\Delta G = \Delta G_{\text{chem.}} + \Delta G_{\text{elec.}} + \underbrace{\Delta G_{\text{ATP synth}}}_{\text{New Term}} + \underbrace{\Delta G_{\text{Maxwell demon}}}_{\text{New Term}}, \quad (2)$$

which, written explicitly, becomes:

$$\Delta G = 2.3 nRT\Delta\text{pH} + n\mathcal{F}Z\Delta\psi + \underbrace{n\mathcal{F}Z\Delta\psi_{\text{ATP synth}}}_{\text{New Term}} + \underbrace{nRT\ln 2}_{\text{New Term}}, \quad (3)$$

The new physics in Eq. (3) is encapsulated in the last two terms with opposite signs. Note that all Δ's in the first three terms are < 0 while the last term is > 0.

ATP synthase is, therefore, more than just a biological catalyst. In addition to its known enzymatic function, it appears to also act as (*i*) a regulator of the kinetics of proton translocation and (*ii*) as a direct participant into the Δ*G* associated with proton translocation.

To help visualize the potential barrier, Fig. 3 presents an averaged projection of the protein's **E** along the *z*-direction and a display of the electric field lines. Clearly, near the entrance of the $F_O$ unit, the electrostatic field is repulsive to positive species, while it generally has the reverse topography near the point of the proton exit from the $F_O$ unit (Fig. 3). The Molecular Operating Environment (MOE)[50] has been used to generate the vector field in this figure.

Table 1 Electric potential differences between exit and entry points of protons (in/out $F_O$ unit), $\Delta\Psi_{\text{ATP synth.}}$, and electric potential barrier, $\Delta\Psi^{\ddagger}_{\text{ATP synth.}}$, and the corresponding Gibbs energies per proton.

| Structure | $\Delta\Psi_{\text{ATP synth.}}$ (mV) | $\Delta G_{\text{ATP synth.}}/H^+$ (kJ/mol) | $\Delta\Psi^{\ddagger}_{\text{ATP synth.}}$ (mV) | $\Delta G^{\ddagger}_{\text{ATP synth.}}/H^+$ (kJ/mol) |
|---|---|---|---|---|
| 5DN6 | -30 | -3 | 21 | 2 |
| 5FL7 | -45 | -5 | 27 | 3 |
| 6CP6 | -8 | -1 | 4 | 0.5 |
| 6J5I | -12 | -1 | 10 | 1 |
| 6N2Y | -18 | -2 | 13 | 1 |

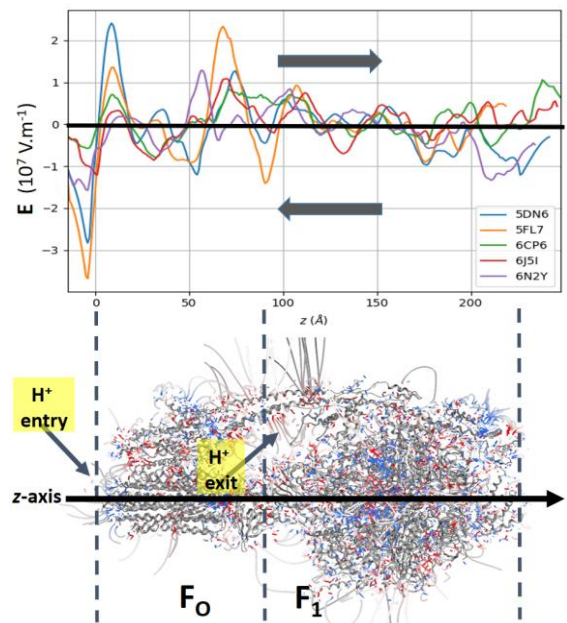

**Fig. 3** (*Top*) The *z*-projection of the intrinsic electrostatic field of ATP synthase, **E** (in $10^7$ V/m), averaged over each plane perpendicular to the *z*-axis between 5 and 15 Å from the protein for the five studied structures. A dark line (**E** = 0) demarcates the top part of the plot (**E** > 0) in which the *z*-projection average points to the right and the bottom (**E** < 0) where it points to the left (indicated by arrows). (*Bottom*) A representation of the electric field lines in the vicinity of ATP synthase (PDB# 6J5I).

At least for the studied systems, evolution seems to have fine-tuned ATP synthase to compensate for the energy it needs to dissipate as it regulates the traffic of $H^+$ by having an intrinsic MESP that "pays *that* bill". The enzyme appears to give with one hand to take it back by the other just to stay alive.

Furthermore, recent examination of the heat conduction around ATP synthase uncovers the role of this enzyme in generating temporary short-lived (picoseconds) temperature gradient spikes with every proton translocation.[51] This view of ATP synthase supports the heat engine character ascribed to this enzyme by Muller.[52;53] Temporal superposition of these spikes in nanothermometry experiments of Chrétien *et al.* can give rise to substantial temperature difference between mitochondria and their surroundings.[54-57] Finally, by driving cell power, ATP synthase determines cellular volume following a scaling law.[58] These observations suggest that ATP synthase is particular and that it is indeed a moonlighting enzyme with non-putative roles.

The authors thank Dr. Nathan Baker (*Pacific Northwest National Laboratory*) and Professor Joelle Pelletier (*Université de Montréal*) for helpful discussions. This research has been supported by the *Natural Sciences and Engineering Research Council of Canada* (NSERC), *Canada Foundation for Innovation* (CFI), *Compute Canada*, *Mount Saint Vincent University*, *Université Laval*, *Saint Mary's University*, and the *Chemical Computing Group, Inc.* J-NV is also grateful to the French *Ministère de l'enseignement supérieur, de la recherche et de l'innovation* for a doctoral scholarship.

*There are no conflicts to declare.*